\documentclass[preprint,proceedings]{rmaa}

\SetYear{2003}
\SetConfTitle{Compact Binaries in the Galaxy and Beyond}

\title{Searching evidences for spiral shocks in the quiescent accretion disk of U Gem} 

\author{
  V. V. Neustroev\altaffilmark{1},
  V. Chavushyan\altaffilmark{2},
  J. R. Vald\'es\altaffilmark{2}
  }

\altaffiltext{1}{National University of Ireland, Galway, Ireland.}
\altaffiltext{2}{Instituto Nacional de Astrof\'{\i}sica Optica y Electr\'onica,
Puebla, Pue., M\'exico.}

\shortauthor{Neustroev et al.}
\shorttitle{Spiral shocks in the quiescent accretion disk of U Gem}

\fulladdresses{
\item V. V. Neustroev: Computational Astrophysics Laboratory, National University 
of Ireland, Galway, Ireland (\protect\email{benj@it.nuigalway.ie}).
\item V. Chavushyan and J. R. Vald\'es: Instituto Nacional de Astrof\'{\i}sica Optica y Electr\'onica,
Apartado Postal 51 y 216, C.P. 72000, Puebla, Pue., M\'exico.
}

\listofauthors{V. V. Neustroev, V. Chavushyan, J. R. Vald\'es}
\indexauthor{Neustroev, V. V.}
\indexauthor{Chavushyan, V.}
\indexauthor{Vald\'es, J. R.}

\abstract{
We find that the quiescent accretion disk of U Gem has a 
complicated structure. Along to the bright spot originating in the 
region of interaction between the stream and the disk particles, there 
are also explicit indications of spiral shocks. The Doppler map 
and the variations of the peak separation of the emission lines are 
indicative.
} 

\addkeyword{Accretion, Accretion Disks}
\addkeyword{Line: Profiles}
\addkeyword{Shock Waves}
\addkeyword{Novae, Cataclysmic Variables}
\addkeyword{Stars: Individual: U Gem}

\begin{document}
\maketitle 

%
Although indications for spiral shocks in the hot accretion disks
during an outburst have already been found \citep{Steeghs97}, 
the problem on the spiral structure
of the quiescent accretion disks still remains unsolved. 
The observational detection of spiral shocks in such
disks would be very important since spiral arms are very efficient 
in transporting angular momentum into the outer part of the disk \citep{Boffin01}.
However if existing, spiral shocks would be much more difficult 
to detect than the strong shocks in the hot
accretion disk during outburst. Nevertheless, first steps towards an
observational confirmation have been started.
\citet{Neus:Bor} and \citet{Neus2002} have found some evidences for spiral
structure of the quiescent accretion disk of U~Gem and IP~Peg.
Further investigations in this area are strongly required.

\begin{figure}[!t]
  \center
  \includegraphics[width=0.83\columnwidth]{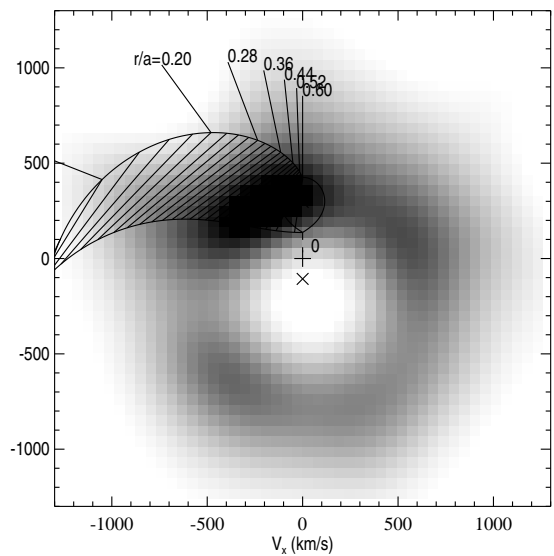}
  \caption{
The Doppler map of the H${\beta }$ emission from U~Gem in quiescence.}
  \label{fig:dop}
\end{figure}

Here we present new evidences for spiral shocks
in the quiescent accretion disk of U Gem.
High signal-to-noise and medium resolution ($\sim$3\AA) optical spectra 
of U Gem were obtained on the 2.12-m telescope of 
the {\it Observatorio Astrofisico''Guillermo Haro''} (AOGH), Cananea, M\'exico, 
during 2000 November 17, with a total coverage of ~4.3 hours.
A total of 24 spectra were taken in the wavelength range 3900-5400~{\AA} with 
exposures of 600 s, covering one orbital period.

The distribution of the accretion disk's emission was explored by computing a
Doppler map, using the method of Doppler tomography \citep{marsh:horne}.
The H$\beta$ tomogram shows
the bright emitting region superposed on the typical ring-shaped emission of 
the accretion disk.
This bright region can be unequivocally contributed
to emission from the bright spot on the outer edge of the accretion disk.
Note that the emission from the secondary is completely absent.

Disk emission is centered on the white dwarf and has
small, but well noticeable azimuthal asymmetry in the form of a two
armed pattern. 
The line flux in the arms is about a factor of $\sim$1.4 stronger than that
of the disk emission outside these areas, pointing to some heating
and density enhancement.
Note that the marked areas of increased luminosity are not perfectly symmetric.
The arm in the upper right of the tomogram is slightly stronger.

\begin{figure}[!t]
  \center
  \includegraphics[width=1.00\columnwidth]{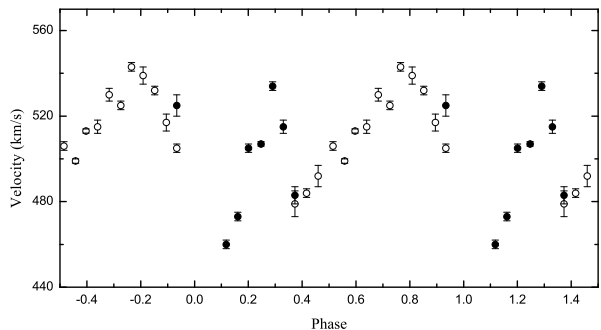}
  \caption{
  The modulation of the double peak separation of the H${\beta }$ emission 
  across the binary period. Filled and open circles correspond to the velocity difference 
  between the center of the line and the blue and red peak respectively. 
  These velocity differences were determined during orbital phases when the emission 
  component from the bright spot was on the opposite peak of the line.
  }
  \label{fig:inter}
\end{figure}

These arms are located too far from the region of interaction
between the stream and the disk particles. 
None of the theories predict the presence of any bright spots here, 
which are connected with such an interaction.
However, exactly in these areas of the Doppler maps there should be spiral
shocks predicted numerically by a number of researchers
\citep{Japan86}.
Furthermore, very similar two-armed structure was detected by \citet{Groot}
in the accretion disk of U~Gem and by \citet{Steeghs97} in IP Peg,
when both systems were in outburst.

However, unlike \citet{Steeghs97} and \citet{Groot}, we cannot confidently
assert that the form of both arms is spiral. The reason of it can be the fact,
that spiral shocks in quiescence should be tightly wrapped \citep{Steeghs99}.
Hence, the areas on the tomograms corresponding to the spirals
will little differ from a ring. In addition this difference will be difficult
for detecting, taking into account low brightness of these shocks.

\begin{figure}[!t]
  \center
  \includegraphics[width=0.90\columnwidth]{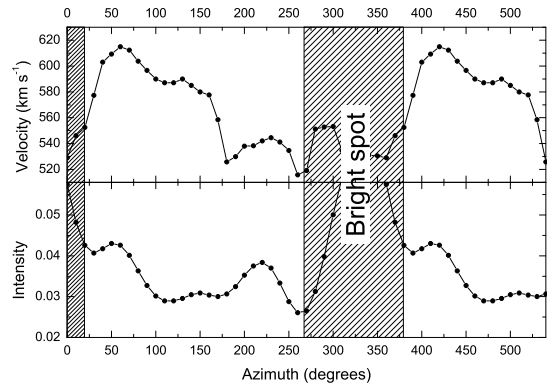}
  \caption{Variation in the velocity and emission strength of the spiral arms 
  as a function of azimuth.}  
  \label{fig:spiral}
\end{figure}

As additional observational evidence for spiral shocks in 
the accretion disk can be regarded
the modulation of the double peak separation of the emission lines across
the binary orbit in a particular way \citep{Neus:Bor}. The detection
of such modulation is complicated by presence of the s-wave component
distorting the lines. To remove this influence, the spectra of U~Gem
were corrected for wavelength shifts due to orbital motion. After that,
we have determined, using gaussian fitting, the radial velocities of only such
line peaks that are not garbled by the s-wave component.
We have taken the absolute values of these velocities as the half of the
double peak separations, and in Fig.~\ref{fig:inter} we show
their dependence on the orbital phase. One can see that the double peak
separation varies during the orbital period, at a first approximation as
$\sin 2\varphi$. These variations are the signature of a m=2 mode in the
accretion disk of U~Gem. This mode can be excited by the tidal forcing and
the detected variations can be explained by the presence of spiral shocks in
the disk, confirming the results of \citet{Neus:Bor}.
Additionally, we have also determined the variation in the velocity and 
emission strength of the (spiral) arms as a function of azimuth 
(Fig.~\ref{fig:spiral}). 
The obtained plot is qualitatively similar to Fig.~4 that has been presented 
by \citet{Harlaftis} for spiral shocks in the accretion disk of IP Peg 
during outburst.

Thus, we find that the quiescent accretion disk of U Gem has a 
complicated structure. Along to the bright spot originating in the 
region of interaction between the stream and the disk particles, there 
are also explicit indications of spiral shock waves. 


\begin{thebibliography}

\bibitem[Boffin(2001)]{Boffin01} Boffin, H.M.J., 2001, Lecture Notes in Physics 573, 69
\bibitem[Groot(2001)]{Groot} Groot, P.J., 2001, \apj, 551, L89
\bibitem[Harlaftis et al.(1999)]{Harlaftis} Harlaftis, E.T. et al., 1999, MNRAS 306, 348
\bibitem[Marsh \& Horne(1988)]{marsh:horne} Marsh, T.R., \& Horne, K., 1988, \mnras, 235, 269
\bibitem[Neustroev \& Borisov(1998)]{Neus:Bor}  Neustroev, V.V., \& Borisov, N.V.,
  1998, A\&A, 336, L73
\bibitem[Neustroev et al.(2002)]{Neus2002}  Neustroev, V.V. et al., 2002, A\&A, 393, 239
\bibitem[Sawada et al.(1986)]{Japan86} Sawada, K., Matsuda, T., \& Hachisu, I., 1986,
        \mnras, 219, 75
\bibitem[Steeghs et al.(1997)]{Steeghs97} Steeghs, D., Harlaftis, E.T., \& Horne, K.,
  1997, MNRAS 290, L28
\bibitem[Steeghs \& Stehle(1999)]{Steeghs99} Steeghs, D., \& Stehle, R., 1999,
  MNRAS 307, 99
\end{thebibliography}
\end{document}